\documentclass[sigconf]{acmart}
\AtBeginDocument{%
  }
\usepackage{graphicx}
\usepackage{amsmath}
\usepackage{hyperref}
\usepackage{subcaption}
\usepackage{xcolor}
\usepackage{url}
\usepackage{float}
\setcopyright{acmlicensed}
\copyrightyear{2025}
\acmYear{2025}
\acmDOI{https://doi.org/10.1145/3773274.3774855}
\acmConference[UCC '25]{18th IEEE/ACM Utility and Cloud Computing (UCC) - Inspire Workshop}{December 01--04, 2025}{Nantes, France}

\acmISBN{978-1-4503-XXXX-X/2018/06}

\begin{document}

\title{Application-level observability \\ for adaptive Edge to Cloud continuum systems}

\author{Kaddour Sidi Mohammed}
\affiliation{%
   \institution{IMT Atlantique, Inria, LS2N}
  \city{Nantes}
  \country{France}
}
\email{sidi-mohammed.kaddour@inria.fr}

\author{Daniel Balouek}
\affiliation{%
   \institution{IMT Atlantique, Inria, LS2N}
  \city{Nantes}
  \country{France}
}
\email{daniel.balouek@inria.fr}

\author{Baptiste Jonglez}
\affiliation{%
   \institution{IMT Atlantique, Inria, LS2N}
  \city{Nantes}
  \country{France}
}
\email{baptiste.jonglez@inria.fr}
\renewcommand{\shortauthors}{Kaddour et al.}
\begin{abstract}
Modern Edge-to-Cloud (E2C) systems require fine-grained observability to ensure adaptive behavior and compliance with performance objectives across heterogeneous and dynamic environments. This work introduces an application-level observability framework that integrates developer-driven instrumentation and SLO-aware feedback for autonomous adaptation. By combining OpenTelemetry, Prometheus, K3s, and Chaos Mesh, the framework enables real-time monitoring and adaptive control across the continuum. A video processing use case demonstrates how application-level metrics guide automatic adjustments to maintain target frame rate, latency, and detection accuracy under variable workloads and injected faults. Preliminary results highlight improved scalability, fault tolerance, and responsiveness, providing a practical foundation for adaptive, SLO-compliant E2C applications.
\end{abstract}


\maketitle

\section{Introduction}

Modern Edge-to-Cloud (E2C) continuum systems face significant challenges in real-time performance management and adaptive operation across heterogeneous and distributed resources.As highlighted in~\cite{Beckman2020HarnessingTC}, the proliferation of billions of interconnected devices across the planet has created a massive, interdependent “system of systems” whose emergent behavior is difficult to predict or control. Traditional monitoring methods, which primarily focus on infrastructure-level metrics such as CPU or memory usage, are insufficient for capturing application-specific behaviors, including error rates, queue lengths, or throughput. Similarly, adaptation strategies often remain limited to coarse-grained infrastructure actions, such as scaling replicas or allocating additional CPU resources, without considering application-level metrics and actions. Fine-grained adaptation, such as adjusting frame rates, reducing processed frames, or switching to lighter inference models, is essential for maintaining performance and quality under varying workloads and constrained conditions.

E2C applications typically consist of multiple microservices, each generating its own metrics and distributed traces. Collecting and analyzing this data at the microservice level enables precise detection of performance anomalies and resource bottlenecks, allowing autonomous feedback mechanisms to make context-aware adaptation decisions.

This capability is particularly important in \textbf{Urgent Computing}~\cite{balouek2020urgent} or \textbf{time-critical computing}~\cite{9210325} scenarios, where services must be deployed and reconfigured rapidly without manual intervention. Examples include disaster response, emergency video analytics, and environmental monitoring, where real-time operation and adherence to performance objectives are critical despite fluctuating resources or network conditions.

Despite this need, there is a lack of open-source use cases enabling experimental evaluation of application-level feedback mechanisms in realistic E2C deployments. Existing work often focuses on simulations or isolated components without providing an end-to-end observable and adaptive pipeline.

To address these gaps, we introduce an \textbf{application-level observability framework for adaptive E2C continuum systems}. Our approach integrates OpenTelemetry, Prometheus, and K3s to collect per-microservice metrics and traces, enabling \textit{multi-layer adaptation} through both infrastructure and application-level actions. The framework supports \textit{metric scalability}, allowing developers to dynamically add, remove, or redefine metrics without modifying the system’s core logic. We validate our approach through a \textbf{video surveillance use case}, demonstrating how application-level metrics guide adaptive decisions to maintain target frame rates, latency, and detection accuracy across heterogeneous and time-sensitive E2C environments.

\section{Background and Related Work}

Traditional monitoring, focused on infrastructure-level metrics such as CPU and memory usage, often fails to provide the application-specific insights required for effective Edge-to-Cloud (E2C) management. This section reviews relevant background concepts and existing solutions, highlighting current gaps.

Ensuring performance objectives in E2C environments is challenging due to distribution, heterogeneous resources, and dynamic conditions. Traditional monitoring often relies on static thresholds and reactive responses, insufficient for real-time demands. Studies emphasize the need for dynamic QoS metrics and adaptive mechanisms to maintain performance~\cite{sauret2025survey}. Research on urgent computing scenarios also highlights adaptive systems capable of rapid deployment and reconfiguration under time-critical conditions, ensuring continuity in disaster response, emergency analytics, or environmental monitoring~\cite{dazzi2024urgent}.
Adaptive systems allow applications to adjust to changing workloads, resource availability, and network conditions. Key approaches include:\\
\textbf{Automatic Performance Diagnosis and Recovery:} Wu~\cite{wu_automatic_nodate} proposes cloud microservice systems that detect anomalies, pinpoint root causes, and recommend recovery actions without explicit instrumentation, offering insights for E2C observability.\\
\textbf{Intent-driven Infrastructure Adaptation:} Proteus~\cite{Proteus2024} translates high-level performance intents into actionable resource configurations for Edge Sensor Nodes, demonstrating autonomous adaptation.\\
\textbf{Decentralized Orchestration:} Markov blanket techniques enable edge devices to infer optimal configurations while preserving QoS for video streams~\cite{monti_designing_2023}. Integrating these into holistic feedback loops remains challenging.\\
\textbf{Proactive Orchestration:} ProKube~\cite{Ali2024}, Tran et al.~\cite{tran2022proactive}, and Khan et al.~\cite{khan2020energy} propose proactive container scaling, migration, and consolidation. Mobile-Kube~\cite{doyle2022mobilekube} and Premsankar~\cite{premsankar2022} address DRL-based migration and DNN placement, highlighting trade-offs between latency, performance, and energy, though often omitting application-level metrics.\\
\textbf{Fault-Tolerant Edge Infrastructure:} Rasouli et al.~\cite{rasouli2024} study MEC infrastructures for mission-critical IoT, using Kubernetes for self-healing and RabbitMQ for reliable messaging, ensuring continuity under node failures or network disruptions.\\
\textbf{Root Cause Analysis (RCA):} Sketch-based anomaly detection~\cite{troya_microsketch_2022}, multi-modal observability~\cite{yu_nezha_2023}, dependency graphs~\cite{hou_diagnosing_2021}, real-time anomaly propagation~\cite{wu_microrca_2020}, and feature reduction~\cite{tsubouchi_metricsifter_2024} support RCA in cloud-edge systems~\cite{zhu_root_2024}. However, RCA is rarely embedded in autonomous feedback loops due to computational complexity and difficulty translating outputs into adaptation actions.

Despite these advances, most approaches focus on infrastructure metrics and lack application-level observability, limiting fine-grained, real-time adaptation. Unified solutions combining multi-layer adaptation, developer-driven instrumentation, and application-level metrics remain scarce, underscoring the need for comprehensive feedback mechanisms in heterogeneous E2C environments.

\section{Observability Driven Feedback Mechanism}

Our approach targets continuum-aware applications, where microservices are distributed across edge and cloud environments. Each component produces traces and metrics that are critical for real-time monitoring and SLO compliance. By instrumenting the application code and selecting relevant metrics, the system continuously observes performance, resource usage, and event-related data.

This observability enables the feedback mechanism to detect potential SLO violations and performance anomalies as they occur. Collected metrics serve as input to automated analysis and decision-making, allowing the system to proactively adjust resources, reconfigure services, or trigger corrective actions. In this way, the approach ensures that distributed applications remain responsive, reliable, and adaptive in dynamic Edge-to-Cloud settings.

The SLO-aware feedback mechanism continuously monitors, detects, and resolves SLO violations in Edge-to-Cloud applications. It consists of two key subprocesses operating in a real-time feedback loop: \textit{Status and Causes Inference} and \textit{Reconfiguration and Resolution}.

Before deployment, 
developers provide a \textbf{descriptive configuration file} defining the application's SLO objectives, relevant metrics, potential application-level actions, and interdependencies among microservices. This file serves as a declarative interface between the application and the feedback controller, enabling the system to identify measurable service-level indicators (SLIs) and their associated Service-Level Objectives (SLOs). It also specifies which adaptation levers—such as modifying frame rates, changing detection models, or adjusting resource allocations—can be applied when SLO degradation is detected. 
\begin{figure*}[ht]
    \centering
    \includegraphics[width=0.6\linewidth]{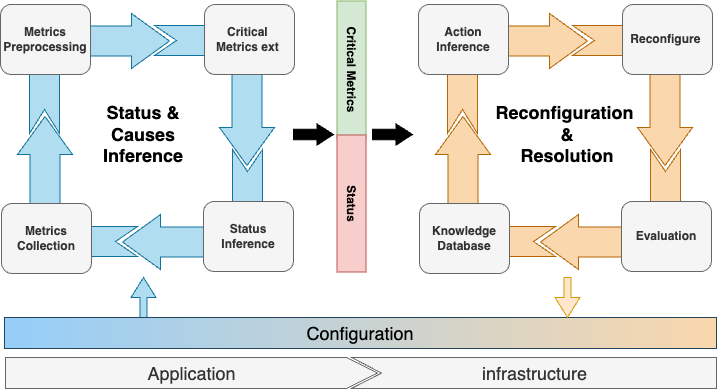}
    \Description{Diagram showing the SLO-aware feedback loop for Edge-to-Cloud applications. The loop includes monitoring of infrastructure and application metrics, analysis to detect SLO violations and identify causes, and reconfiguration actions such as scaling resources or adjusting application parameters. Developers provide a descriptor file specifying relevant SLOs, metrics, actions, and microservice dependencies.}
    \caption{Feedback Mechanism Scheme}
    \label{fig:feedback-mechanism-scheme}
\end{figure*}
By integrating this developer-defined knowledge into the feedback loop, the mechanism ensures that monitoring and adaptation remain application-aware, context-sensitive, and aligned with the intended quality and performance objectives of the system.

\subsection{Status and Causes Inference}

This subprocess monitors application performance across edge and cloud components and detects SLO violations based on predefined thresholds. When a violation occurs, the system analyzes metrics such as processing times, response times, and resource utilization to infer root causes.

\subsubsection{\textbf{Metrics Collection}}
Performance metrics are collected over time \(t\) for each component \(c\) from a Time-Series Database (TSDB):
\[
M_c(t) = f(c, t)
\]
where \(f(c, t)\) returns the value of a specific metric. A dependency graph \(G = (V, E)\) is constructed, where \(V\) represents system components (hosts, pods, services, metrics) and \(E\) represents interactions. Dependencies are quantified as correlations:
\[
R(v_1, v_2) = \mathrm{correlation}(M_{v_1}, M_{v_2})
\]

\subsubsection{\textbf{Metrics Preprocessing}}
Collected metrics undergo standard preprocessing:
\begin{itemize}
    \item Cleaning and Interpolation: Handle missing values and outliers.
    \item Normalization: Min-Max scaling ensures metrics are comparable.
\end{itemize}
\subsubsection{\textbf{Critical Metrics Extraction}}
Critical metrics that impact SLO compliance, such as response times, error rates, and throughput, are identified using anomaly detection methods \cite{liu2008isolation}. The system leverages the dependency graph to prioritize metrics and assess correlations between anomalies, allowing a focused analysis of the components most likely to affect overall performance.

\subsubsection{\textbf{Status Inference}}
The system evaluates metric conditions \(C = (m, \operatorname{op}, v)\) to determine overall status:
\[
\text{status} = 
\begin{cases} 
\text{good} & \text{if all conditions are satisfied} \\
\text{fail} & \text{otherwise}
\end{cases}
\]
This allows real-time detection of SLO violations and identification of components likely responsible.

\subsection{Reconfiguration and Resolution}
Upon identifying the root cause, the system autonomously adjusts parameters to restore SLO compliance. Actions may include resource scaling, load redistribution, workflow optimization, or alternative processing paths. These adjustments ensure system responsiveness, maintain service quality, and operate effectively in time-critical, urgent computing contexts with minimal human intervention.

\subsubsection{\textbf{Action Inference}}
Based on inferred status and root causes, appropriate actions are determined. These actions are categorized into two main types: infrastructure-level (e.g., scaling K3s pods via Kubernetes HPA, adjusting resource limits) and application-level (e.g., dynamically lowering video frame rates, switching to a lighter object detection model, or modifying processing queues). The inference is guided by a predefined set of actions within the descriptive configuration file, which maps specific SLO violations and root causes to a prioritized list of corrective measures.

\subsubsection{\textbf{Reconfiguration}}
After actions are inferred, they are applied to the system. K3s is utilized to reconfigure application or infrastructure components according to the inferred actions. For instance, if an SLO violation is traced to high CPU usage on an edge node, K3s might scale out the affected microservice. If the issue is application-specific, such as high response time in video processing, the system might command the video processing microservice to reduce its frame sampling rate.

\subsubsection{\textbf{Evaluation}}
Following reconfiguration, the impact of applied actions on critical metrics and SLO compliance is continuously monitored. This evaluation ensures that the corrective measures are effective and identifies any remaining or emerging performance degradations.

\subsubsection{\textbf{Knowledge Database}}
To enhance adaptive capabilities, a knowledge database stores historical data on SLO violations, their inferred root causes, and the effectiveness of applied actions. This repository informs future adaptations, allowing the system to learn optimal responses over time and refine its decision-making process through reinforcement learning techniques (planned for future work).
\subsection{Novelty in Application-Specific Metrics and Developer-Driven Instrumentation}

Our approach emphasizes \textbf{application-specific metrics} and \textbf{developer-driven instrumentation}. Unlike traditional monitoring, which relies on generic infrastructure metrics, developers define and instrument metrics reflecting functional and nonfunctional requirements. By integrating OpenTelemetry from the development phase, telemetry data is context-rich, enabling precise root cause analysis and targeted optimization.
\section{Evaluation}
\subsection{Use Case: Video Processing Application}
\begin{figure}[h]
    \centering
    \includegraphics[width=1\linewidth]{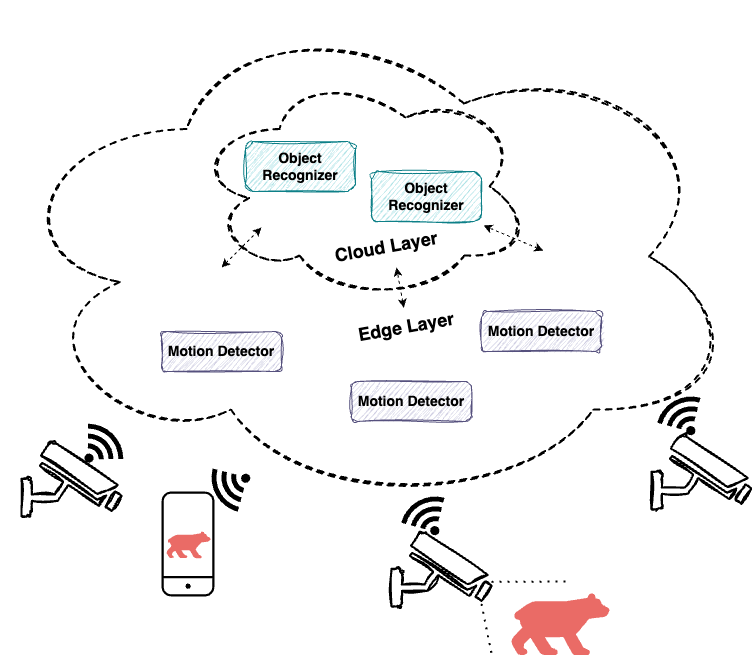}
    \Description{Overview of the Edge-to-Cloud use case showing video frames captured at the edge, motion detection performed on edge nodes, and relevant frames sent to cloud-based object recognition. The diagram highlights the flow of metrics and SLO-aware feedback actions across microservices.}
    \caption{Overview of the Edge-to-Cloud Use-Case}
    \label{fig:use-case}
\end{figure}
The surveillance system ,previously introduced in our earlier work \cite{dais2025},is a distributed Edge-to-Cloud architecture designed to process video feeds and identify dangerous animals, ensuring timely alerts for residents and authorities. It comprises the following key components,

\textbf{Camera}: Captures video frames, resizes them for efficient processing, and transmits them to the motion detection module.

\textbf{Motion Detection}: Operates at the edge, analyzing video frames for motion and forwarding relevant frames to the object recognition module.

\textbf{Object Recognizer}: Uses the YOLO model for object classification and detection. It operates in the cloud and processes frames asynchronously.\\
Figure~\ref{fig:use-case} shows the flow from edge cameras to cloud object recognition.

The application leverages OpenTelemetry to collect and send application-level metrics for the feedback mechanism.

This application is open-source and available at \url{https://gitlab.inria.fr/STACK-RESEARCH-GROUP/software/edge-to-cloud-video-processing.git}.

\subsection{Setup}
\begin{figure}[h]
    \centering
    \includegraphics[width=1\linewidth]{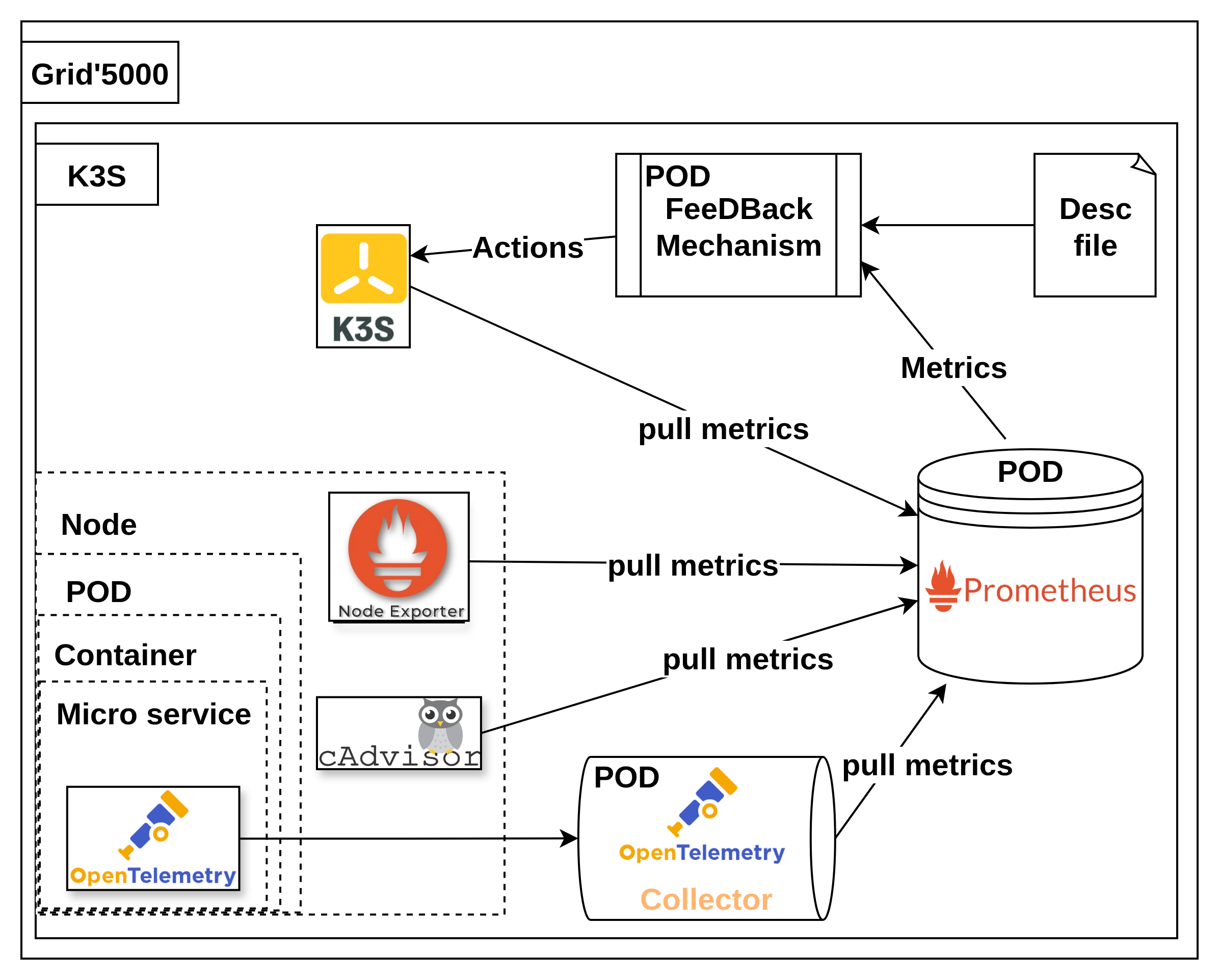}
    \Description{Observability workflow for the Edge-to-Cloud use-case on Grid\'5000. Shows metrics and traces collection, analysis, and adaptive actions.}
    \caption{Observability Workflow for the Use-Case Application on Grid\'5000}
    \label{fig:observability-workflow}
\end{figure}

We deployed a preliminary experimental environment on Grid'5000 \cite{balouek2012adding} to evaluate our SLO-aware feedback mechanism. The setup uses several nodes (G5K virtual machines) with K3s (via EnOSlib) simulating camera, edge, and cloud environments. Camera nodes are lightweight devices with 1 CPU core and 1 GB RAM, responsible for generating and transmitting video frames. Edge nodes have 4 CPU cores, 8 GB RAM, and 1 Gbps connectivity, while cloud nodes have 8 CPU cores, 16 GB RAM, and 10 Gbps connectivity. OpenTelemetry collects application traces and custom metrics (e.g., frame processing time, detection accuracy), exported to Prometheus. K3s tools (cAdvisor, Node Exporter) monitor CPU, memory, and network usage. Chaos Mesh injects faults (e.g., network latency, CPU pressure) to test resilience and adaptation.
\begin{table}[h]
\centering
\caption{Grid'5000 Experimental Hardware Setup}
\label{tab:hardware-setup}
\begin{tabular}{|l|l|}
\hline
\textbf{Component} & \textbf{Specs} \\ \hline
Camera Nodes & 1 CPU, 1 GB RAM \\ \hline
Edge Nodes & 4 CPU, 8 GB RAM \\ \hline
Cloud Nodes & 8 CPU, 16 GB RAM \\ \hline
\end{tabular}
\end{table}

\begin{table}[h]
\centering
\caption{Metrics Collected by Tools and Their Nature}
\label{tab:metrics-tools}
\begin{tabular}{|l|l|}
\hline
\textbf{Tool} & \textbf{Metric level} \\ \hline
OpenTelemetry & Application (frame rate, processing time, accuracy) \\ \hline
cAdvisor & Infrastructure (CPU usage,  container stats) \\ \hline
Node Exporter & Infrastructure (CPU load, memory, network I/O) \\ \hline
\end{tabular}
\end{table}

\subsection{Results}
We present results from two complementary evaluations: one assessing system scalability as the number of cameras and motion detectors increases, and another examining the impact of application-level metrics, such as animal appearance rates, on system response times. These experiments illustrate the system’s behavior under varying workloads and metric conditions.

\begin{figure*}[ht!]
    \centering
    \includegraphics[width=0.6\linewidth]{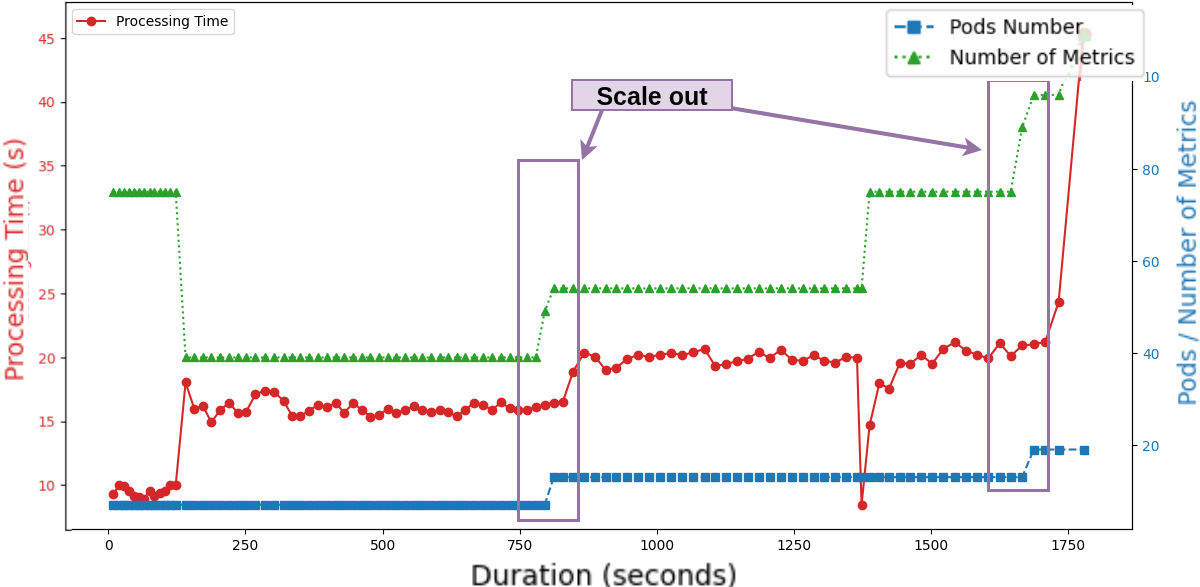}
    \Description{Graph showing the system adaptation over time, including metrics collected from microservices, processing time variations, and the number of pods deployed. Demonstrates how the SLO-aware feedback loop adjusts infrastructure and application parameters to maintain performance under varying workloads.}
    \caption{System evolution over time showing processing time, metrics, and pod count; the scale-out window indicates adaptive resource adjustment under load.}
    \label{fig:system-adaptation}
\end{figure*}

    \subsubsection{\textbf{Scalability}} 
Figure~\ref{fig:system-adaptation} illustrates that our feedback mechanism adapts seamlessly to system growth and evolving monitoring needs. When the video surveillance system scales out by adding three cameras and three motion detectors, the mechanism dynamically adjusts the allocation of resources, maintaining stable processing times. The strong correlation observed between metrics, processing time, and pod scaling confirms efficient adaptation and resource utilization. Furthermore, our approach supports \textit{metric scalability}: developers can add, remove, or modify monitored metrics at runtime without disrupting operation or requiring code changes. This dual scalability—both at the system and metric levels—demonstrates the robustness, flexibility, and practicality of our feedback mechanism for real-world Edge-to-Cloud monitoring scenarios.

\subsubsection{\textbf{Applicative Metrics Inclusion}} 

Applicative metrics provide valuable insights into system behavior, enabling optimizations beyond infrastructure-level monitoring. In this context, we define the \texttt{APPEARANCE\_RATE (AR)} as the expected number of times an animal appears per hour in a simulated footage. The response time of the object recognizer is directly influenced by the number of detected motions.
\begin{figure}[H]
    \centering
   \includegraphics[width=1\linewidth]{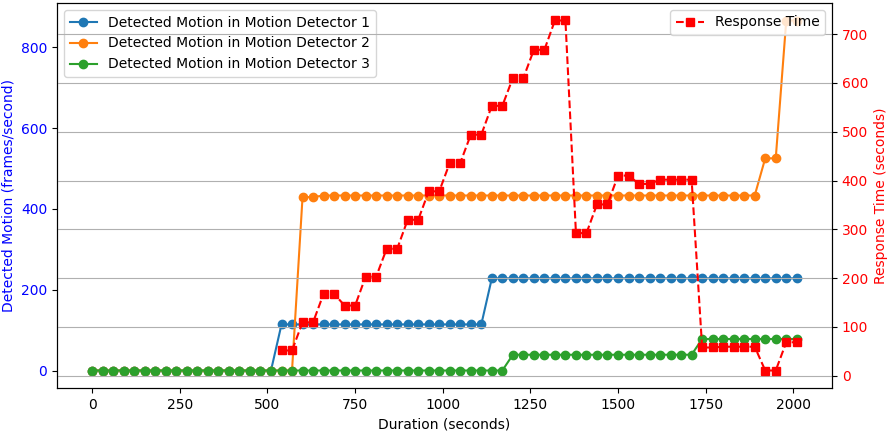}
    \Description{Detected motions versus response time with appearance rates 05, 05, 05. Demonstrates system responsiveness under low motion frequency.}
    \caption{Detected motions vs response time (AR: 05, 05, 05)}
    \label{fig:subfig1}
\end{figure}
At low appearance rates (AR: 05, 05, 05), as shown in Figure~\ref{fig:subfig1}, the system maintains stable response times, indicating efficient processing under light workloads.

\begin{figure}[H]
    \centering
   
    \includegraphics[width=1\linewidth]{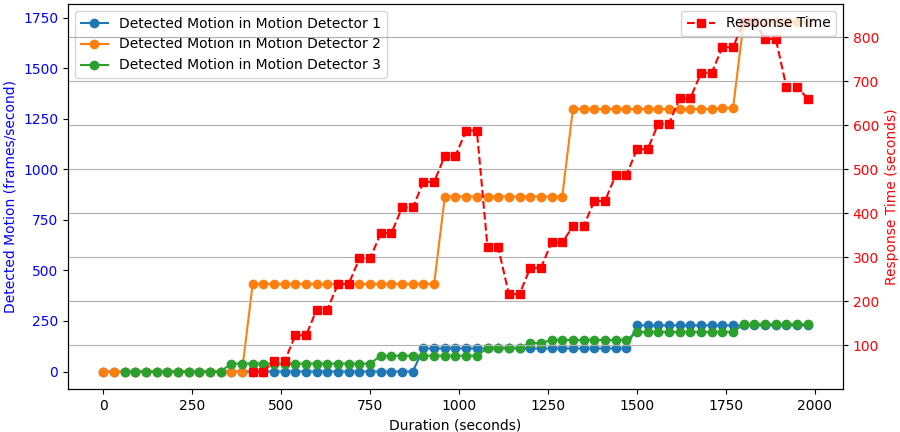}
    \Description{Detected motions versus response time with appearance rates 05, 10, 15. Demonstrates system response under mixed motion frequencies.}
    \caption{Detected motions vs response time (AR: 05, 10, 15)}
    \label{fig:subfig2}
\end{figure}

When the motion frequency increases and becomes unbalanced (AR: 05, 10, 15), as shown in Figure~\ref{fig:subfig2}, response times vary according to the motion density in each camera stream, revealing the sensitivity of the object recognizer to fluctuating workloads.

\begin{figure}[H]
    \centering
    
    \includegraphics[width=1\linewidth]{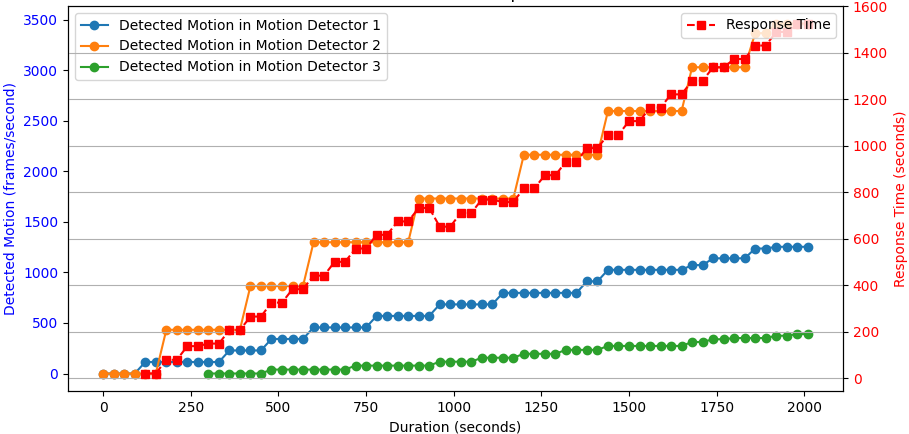}
    \Description{Detected motions versus response time with appearance rates 30, 30, 30. Demonstrates system behavior under high motion frequency.}
    \caption{Detected motions vs response time (AR: 30, 30, 30)}
    \label{fig:subfig3}
\end{figure}

Finally, under high appearance rates (AR: 30, 30, 30) shown in Figure~\ref{fig:subfig3}, the response time increases significantly due to the higher number of detected motions. In such conditions, the system can dynamically reduce the sampling rate to maintain stability and prevent overload.

\section{Discussion}

This work demonstrates the potential of application-level observability combined with performance-aware feedback for adaptive Edge-to-Cloud (E2C) systems. By leveraging developer-driven instrumentation, OpenTelemetry, Prometheus, K3s, and Chaos Mesh, the proposed framework enables real-time monitoring and autonomous adaptation across heterogeneous and dynamic environments. 

The preliminary results from the video processing use case illustrate how application-specific metrics, such as frame processing time and object appearance rates, guide adaptive actions. These actions include adjusting the frame rate, reducing the number of processed frames, or switching to a lighter detection model, ensuring that target performance objectives—such as latency, frame rate, and detection accuracy—are maintained even under variable workloads and injected faults. 

The framework’s support for multi-layer adaptation, encompassing both infrastructure and application-level actions, highlights its capacity to maintain compliance with defined performance objectives while efficiently utilizing available resources. Furthermore, the ability to dynamically add, remove, or redefine metrics without code changes demonstrates the flexibility and scalability of the approach in evolving E2C environments.

Overall, the study provides evidence that fine-grained, application-level observability enables more precise and context-aware adaptation than traditional infrastructure-centric monitoring. These insights lay a practical foundation for developing resilient, performance-compliant, and self-adaptive E2C applications.

\subsection{Future Directions}
Our future work revolves around enhancing the automated configuration of microservice dependencies and adaptation rules for E2C applications. Future work will explore SLO enforcement through three key directions: (i) a comprehensive experimental validation for scalability and SLO compliance, (ii) machine learning–driven adaptation to predict SLO violations, and (iii) automated metric discovery and management.

\section{Conclusion}
In this paper, we introduced an application-level observability framework for adaptive Edge-to-Cloud (E2C) continuum systems. The approach leverages developer driven instrumentation and  application-specific metrics to enable real-time monitoring, analysis, and self-adaptation across heterogeneous environments. By combining tools such as OpenTelemetry, Prometheus, K3s, and Chaos Mesh, the system achieves continuous observability and responsive adaptation.

The video processing use case demonstrated how application-level insights improve resilience, scalability, and responsiveness under dynamic workloads and fault conditions. This work establishes a foundation for autonomously managed continuum applications and paves the way toward intelligent, self-optimizing systems that reduce operational complexity while enhancing reliability and performance.
\begin{acks}
This research was supported by two projects: the French project \textbf{OTPaaS} funded by Bpifrance, and the \textbf{QUICK - Urgent Computing across the Edge-Cloud Continuum} project, funded by the Etoiles Montantes program - Région Pays de la Loire, France.
\end{acks}
\bibliographystyle{ACM-Reference-Format}
\bibliography{bibliography}

\end{document}